# PULP: Inner-process Isolation based on the Program Counter and Data Memory Address

Xiaojing Zhu[1], Mingyu Chen[1], Yangyang Zhao[1,2], Zonghui Hong[1,2], Yunge Guo[1,2]

[1]Institute of Computing Technology, Chinese Academy of Sciences

[2]University of Chinese Academy of Sciences

**Abstract**

Plenty of in-process vulnerabilities are blamed on various out of bound memory accesses. Previous prevention methods are mainly based on software checking associated with performance overhead, while traditional hardware protection mechanisms only work for inter-process memory accesses. In this paper we propose a novel hardware based in-process isolation system called PULP (**P**rotection by **U**ser **L**evel **P**artition). PULP modifies processor core by associating program counter and virtual memory address to achieve in-process data isolation.

PULP partitions the program into two distinct parts, one is reliable, called primary functions, and the other is unreliable, called secondary functions, the accessible memory range of which can be configured via APIs. PULP automatically checks the memory bound when executing load/store operations in secondary functions. A RISC-V based FPGA prototype is implemented and functional test shows that PULP can effectively prevent in-process bug, including the Heartbleed and other buffer overflow vulnerabilities, etc.

The total runtime overhead of PULP is negligible, as there is no extra runtime overhead besides configuring the API. We run SPEC2006 to evaluate the average performance, considering the LIBC functions as secondary functions. Experimental timing results show that, running bzip2, mcf, and libquantum, PULP bears low runtime overhead (less than 0.1%).

Analysis also shows that PULP can be used effectively to prevent the newest "Spectre" bug which threats nearly all out-of-order processors.

## 1 Introduction

Many scenarios could induce in-process vulnerabilities, leading to information leakage and control flow corruption. Applications always call third-party library or components, which are not controlled by programmer and may have security vulnerabilities. When the calling function transmits parameters to the called function, the latter cannot ascertain the expected range of the parameters, buffer overflow may be brought in and the control flow may be deviated. By injecting malformed codes into a victim program's address space or using ROP like techniques, attackers could steal sensitive data from victim program.

Recently, "Spectre" exploit critical vulnerabilities in modern Out-of-Order processors to bypass software boundary check. Spectre[1] involves victim to speculatively perform operations that would not occur during correct program execution and may probably leak the victim's confidential information via a side channel to the adversary.

There have been various methods to mitigate inner-process vulnerabilities, such as MPX etc. However, they are mostly based on software checking or software enforced range checking. The drawback is, 1. need modify third party code. 2 software-overhead is large. 3. can be bypassed by Spectre.

The main cause of in-process vulnerabilities above is that modern processors lack hardware mechanisms to check inner-process memory accesses. Traditional memory protection is based on process address space defined by operating system. Only memory accesses across process domains are checked and prevented if violations are detected. The memory access boundary checking within a process are accomplished by software, which is neither complete nor efficient due the complexity of software.

If we only allow the reliable program codes to access sensitive critical data, we can prevent malicious or buggy untrusted code from inspecting the critical data in advance.

In this paper we propose a novel hardware based in-process isolation system called PULP. PULP modifies the processor pipeline, especially the implementation of load/store instructions, to check whether the inner-memory access address is legal.

PULP separates user process into trusted and untrusted parts, limiting the memory access ranges of untrusted part. PULP inserts the API function before the function calling, which can restrict the data ranges that the called function can access. To separate the trusted functions from untrusted functions, PULP needs also to modify the OS process loader slightly.

When executing the load/store instruction, PULP don't need additional instructions to check the memory address. The software runtime overhead of PULP is negligible compared to Intel MPX.

We use a RISC-V Rocket-chip platform to implement our prototype. We add new registers to the CPU core, and modify Linux kernel to support register configuration and new out of memory bound exception handler. The CPU frequency of RISC-V remains 62.5 MHz. The hardware comparison results show that, the area of PULP is bigger than the old rocket chip version by 31%, the cells of PULP is more than rocket chip by 2%, the power of PULP is more than rocket chip by 28%.

We choose MIT benchmarks[2] and Heartbleed[3] as our security test benchmarks. The benchmark sources are modified to add range-adjustment APIs.

Experimental results show that PULP can effectively prevent buffer overflow and memory leakage vulnerabilities ahead of time.

We also use SPEC2006 to test the performance of PULP. As the load/store instructions of PULP automatically check the memory access address at runtime, no additional instruction is needed to check the memory address. Results show that PULP incurs low runtime overhead (less than 0.1%) to SPEC2006.

Since Rocket chip is an in-order processor, we cannot reproduce Spectre attack, but the in-process data isolation mechanism of PULP can effectively prevent Spectre, and prohibit the unreliable part of the program from gaining the security sensitive data whether it is invoked by speculation or not.

With PULP we contribute the following:

- We propose a novel inner-process hardware based memory protection mechanism PULP. PULP provides APIs to prevent the secondary functions from accessing critical inner-process memory region.
- We have implemented a prototype system with an enhanced RISC-V CPU core, modified Linux kernel as well as a set of APIs to demonstrate PULP.
- We have testified the PULP prototype by both functional and performance benchmarks. The result shows that PULP can effectively defend buffer overflow and Heartbleed problems.

The rest of the paper is organized as following: in section 2 we lay out the motivation of PULP. Section 3 gives the design and implementation of PULP. Section 4 provides the experimental evaluation results. We contrast the related work in section 5 and conclude the paper in section 6.

## 2 Motivation

### 2.1 Inner-process Abuse

Inner-process memory abuse ranges from data theft to privilege escalation. Various of user-space attacks can succeed once they have penetrated into targets' process context[8], as they can freely access (or abuse) target programs' memory content.

Inner-process abuse is usually caused by improper internal memory access, and triggered by many unreliable factors inside user process, such as third-party program, improper function parameters, buffer overflow and speculation, etc. Buffer overflow or out of bound memory access are the main source of inner-process abuse.

Buffers are areas of memory set aside to hold data transited from one function to another. Malformed inputs like an anomalous transaction that produces more data than expected could lead to writes beyond the end of the buffer, which is called Buffer overflows[12]. If the overflow overwrites adjacent data or executable code, it will probably result in erratic program behavior, including memory access errors, incorrect results, and crashes.

## 2.2 Limitation of existing methods

There are many classic software countermeasures against Buffer Overflow, such as Libsafe[14], LibsafeXP[4], stack protector[5], stack canaries[17], etc. These classic software methods always have much large overhead. Though the stack protector has little runtime overhead, it can only protect variables in the stack.

Recently there are some hardware and software combined methods to achieve inner-process isolation, such as Intel MPX[6], Dune[7] and Shreds[8]. Among them, Intel MPX has high runtime overhead, and because of cross privilege, Dune has difficulty when sharing data between the trusted and untrusted parts of the process, while Shreds has higher overhead than PULP.

## 2.3 Main idea of PULP

The main difficulty for inner-process protection is lack of a hardware mechanism to identify different address space within a Process. Previously hardware assumes all code can access the address space of the whole process while leave the software to protect itself. Instead of adding complex hardware domains within a process, PULP chooses to use the address of instruction code to identify a subject. By associating the address of instruction with the data address to be accessed, a protection check can be enforced by hardware with minimum change to the software code.

## 3 System design and implementation

Depending on the reliability of each part of process codes, we divided the code region of the process into many segments. In this paper, each code segment is a function. Before an untrusted function is called, PULP binds the function with its accessible memory regions by configuring the accessible memory regions in special registers. These new registers associate function with specific accessible data regions.

When user process is executed, CPU will check if the data access issued by specific function is legal according to its bounded data regions indicated by registers. Any out of range access attempt will be prevented by CPU pipeline.

To realize this, PULP needs hardware modification and software supplement. The hardware modification includes newly added registers and processor pipeline modification to limit the scope of both load/store instructions and control flow direction.

The software supplement includes three parts, API functions, compiler and kernel. The APIs are designed to define the accessible range of a specific function. The compiler decides which function is the primary function, and the data memory ranges the secondary function can access. The kernel is modified to support the new registers and the new added out of bound memory exception.

As shown in Figure 1, compiler verifies PC range of primary function and safe areas of secondary function's input parameters, instruments program with inline API. Then the kernel configures the registers. Finally the CPU executes the program under specific memory access limitation.

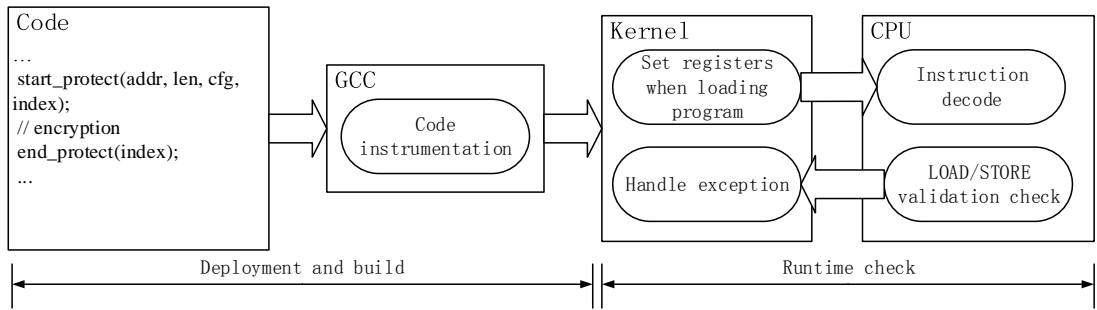

**Fig. 1: Developers can instrument programs via APIs; Compiler can also automatically analyze, detect the primary function range, instrument programs with APIs; during startup, Kernel configures the registers; when executing program, CPU limit access ranges of data memory access.**

### 3.1 Hardware design

#### 3.1.1 new registers

PULP added three kinds of registers in the CPU core, PPCR, SMAR, RAR, which are used to restrict the accessible memory ranges of the untrusted code.

There is one group of primary program counter range (PPCR) registers, consist of the starting and ending program counter address of the primary function. There are several groups of secondary memory address range (SMAR) registers, each group including one pair of starting and ending address of code range. Whenever CPU executes load/store instruction, the pipeline looks up these registers, confirming whether the load/store instruction is legal.

There are two rules about the PPCR and SMAR registers:

Rule 1. The primary function could access its local variables and all the global variables, while the secondary function could only access ranges the SMAR registers indicated.

This rule prevents the secondary functions accessing sensitive data in the primary function.

Rule 2. Only kernel can modify PPCR. Only primary function could modify SMAR registers.

This rule forbids secondary function to configure SMAR registers, which assures that unreliable code could not change its accessible memory ranges.

To avoid ROP-like attack, PULP add a return address register(RAR) in the CPU core, recording the return address of secondary function. When the secondary function is called, the return address will be stored into RAR. When the secondary function returns, the target PC will be compared with PC stored in RAR, and mismatch of PC will trigger return-address-error exception.

With the help of RAR, we can assure the control flow integrity.

#### 3.1.2 processor pipeline

In order to realize the address cross-border judgement, a series of address comparisons are added in the pipeline. If the memory access address or the branch target is beyond the specified range, a corresponding abnormal signal is generated. We take a typical five-stage pipeline as example.

**IF** (Instruction fetch). An instruction is fetched from the memory by program counter.

**ID** (Instruction decode). If the decoded instruction is LOAD, STORE, BRANCH or JUMP, PULP will judge which region the instruction belongs to, such as kernel, primary function or secondary function.

If in kernel mode, no checking is needed. If in user mode, and the PC is within the address range of primary function as set in PPCR, no further checking is needed.

Otherwise PC is in the secondary function, checking will be needed in the EX stage.

**EX** (Execution). According to the results of the ID stage, if the instruction is LOAD/STORE and it belongs to secondary function, the memory address will be checked, and PULP judges whether it is within legal memory bounds set in SMAR.

If the instruction jumps from primary to secondary function, the return address will be stored into RAR.

If the instruction returns from secondary function to primary function, PULP compares target address with return address stored in the RAR. If they are the same, this return is correct. Otherwise this return is wrong, PULP will trigger a return address error exception.

**WB** (Write back). During this stage, the data loaded from memory or calculated by the ALU would be written to the register file.

### 3.2 Software design

#### 3.2.1 PULP API

Programmers can use APIs below conveniently to deploy PULP:

s*tart_protect (addr, len,cfg,index);*
*end_protect (index);*

*start_protect* writes the *addr* into the lower one of SMAR registers' pair, and the sum of *addr* and *len* into the upper one. The *cfg* defines the secondary function's permission of the address set in SMAR.

When running programs, PULP will inquire the SMAR registers to see if the following operations are within the specific memory. In this way, malicious access to memory space would never succeed as they will be prevented by PULP, and raise exception.

The index is needed to indicate specific registers used by the current API.

After invocation of secondary function, *end_protect* API is invoked to clear a SMAR group specified by index, so that PULP will not check the memory access range anymore.

To prevent secondary functions modifying SMAR registers, special configuration instructions in *start_protect* and *end_protect* could only be executed in primary function.

#### 3.2.2 Compiler modification

The compiler computes input parameter's effective data length, and instruments *start_protect* and *end_protect* APIs before and after the called site of the secondary function.

The compiler also figures out which part of the program is primary function, that is, the most trustable part of the program, which will guide the kernel to fill the values of PPCR registers.

Moreover, definition for the new instruction to manipulate the new registers is added to the compiler.

#### 3.2.3 Kernel modification

Loadelf function in the kernel is modified, to configure the PPCR registers. Context switching code of kernel is modified, to support the new registers. And the new exception handler is added, to handle the out of memory bound exception and return-address-error exception.

PPCR registers are set by kernel. When loading elf file, Linux gets the primary function's PC ranges in the *load_binary* function, and set them into the PPCR registers.

Furthermore, we added the definition of PULP registers in kernel, and updated context-switching codes.

## 4 Analysis and evaluation

RISC-V[19,20] is a free and open source instruction set architecture (ISA) based on modern design techniques and decades of computer architecture research. Rocket Chip is an open source RISC-V system-on-chip design generator, which is highly synthesized and capable of generating RTL (Resistor Transistor Logic). To limit the data access ranges and control flow direction, we modified the RISC-V processor pipeline based on Rocket Chip.

All the experiments were carried out on the modified RISC-V processor with 1G RAM memory running Linux kernel 4.6.2, the processor basic frequency is 62.5Mhz.

## 4.1 Security Evaluation

### 4.1.1 Heartbleed attack

Heartbleed is a security bug in the OpenSSL cryptography library, which is a widely used implementation of the Transport Layer Security (TLS) protocol. It was introduced into the software in 2012 and publicly disclosed in April 2014. It results from improper input validation (due to a missing bounds check) in the implementation of the TLS heartbeat extension[13], thus the bug's name derives from heartbeat[3].

Heartbleed originates in buffer overread vulnerabilities, which means reading more data than the software approved[16]. The Heartbleed vulnerability took the Internet by surprise in April 2014. The vulnerability allowed attackers to remotely read protected memory from an estimated 24–55% of popular HTTPS sites.

We implemented a socket server with the API provided by OpenSSL-1.0.1e in C, which initializes the socket, SSL library, and waits for the connection of client using *SSL_accept*.

We wrote a socket client program in C, to communicate with socket server. The client sends hello request of TLSv1.1 to server, and the connection between server and client will be established.

To replay Heartbleed, we sent a malformed Heartbeat request in client. As expected, the server replies with excess data that may be secret.

After deploying PULP in OpenSSL source code, that is, putting the *memcpy* invocation in *tls1_process_heartbeat* under protection, the unexpected copy operation raised an exception and the process is terminated by kernel. Finally no secret data is leaked out.

### 4.1.2 MIT benchmarks

We also used MIT Lincoln Laboratories buffer overflow benchmark [2]. It contains model programs with and without buffer overflow bugs developed from reported vulnerabilities in three real-world open source network servers, namely Bind, Wu-ftpd and Sendmail. Each of the three application programs has several bugs reported in CVE and CERT data bases and accordingly captured in the model programs.

We test 6 cases of MIT benchmark, apply PULP to every bad memory operation that may cause corruption. Then we test them on platform RISC-V without and with PULP.

Without PULP, these 6 test cases cause memory corruption, for example, information leakage, and local variables coverage. After deploying PULP, all of the illegal operations are prevented and raise exception.

These test cases are got from SARD (Software Assurance Reference Dataset)[18]. Their test case id numbers are 1283, 1285, 1289, 1291, 1295, 1297 respectively.

1283 is Off-by-one overflow from MIT benchmarks models/wu-ftpd/f2. 1285 is Realpath() overflow from MIT benchmarks models/wu-ftpd/f3. 1289 is nslookup Complain vulnerability from MIT benchmarks models/bind/b4. 1291 is SIG-BUG from MIT benchmarks models/bind/b2. 1295 is IQUERY-BUG from MIT benchmarks models/bind/b3. 1297 is Remote Sendmail Header Processing Vulnerability from MIT benchmarks models/sendmail/s1.

## 4.2 Performance Evaluation

### 4.2.1 GCC stack protector

To evaluate the API configuration overhead of PULP, we program two micro benchmarks, both of which execute *strcpy* 10000 times, and the input string length of strcpy is 100.

To compare performance of PULP and stack protector, one program is configured with the PULP

API. And another similar program is compiled with GCC option -fstack-protector.

In Figure 2 we compare execution times of PULP and stack protector. The former is 0.29s, the latter is 0.26s. The execution time of stack protector is less. Which is almost the same as original execution time without any runtime protection. So the configuration time of PULP is less than 15% of the execution time of *strcpy* function, with the input string length is 100.

The total configuration time of PULP is in proportion to the called times of secondary function and the numbers of function parameters. If the called times of *strcpy* are fewer, the input string length of strcpy is longer, the configuration overhead in the total execution time will be relatively lower.

```
# time ./strcpy_micro_stack
test
strcpy done
real    0m 0.26s
useer   0m 0.08s
sys     0m 0.11s
# time ./strcpy_micro_pulp -pulp
test
strcpy done
check is 0
real    0m 0.29s
user    0m 0.11s
sys     0m 0.10s
```

**Fig 2. performance test results of the stack protector and the PULP.**

GCC stack protector need low time overhead. However, GCC stack protector can protect only stack variables, while PULP can protect heap variables and global variables as well. Table 2 and table 3 give two functional test programs of stack protector. The first micro benchmark protects variable in the stack. The second one protects variable in the heap.

**Table 2 functional test program of stack protector, with the protected variable in the stack.**

```
#include "stdio.h"
#include "string.h"
int main(int argc, char *argv[])
{
    int check = 0;
    char pass[10];
    printf("test\n");
    strcpy(pass, argv[1]);
    printf("strcpy done\n");
    return 0;
}
```

**Table 3 functional test program of stack protector, with the protected variable in the heap.**

```
#include "stdio.h"
#include "string.h"
int main(int argc, char *argv[])
{
    int check = 0;
    char *pass;
    printf("test\n");
    pass = malloc(10);
    strcpy(pass, argv[1]);
    printf("strcpy done\n");
    return 0;
}
```

The above programs' execution results are shown in Figure 3 and Figure 4. In Figure 3 stack protector protects stack variable successfully. In Figure 4 stack protector cannot protect variables in the heap. Moreover, GCC stack protector can only check stack variable after the function execution finished. The checkout time may be too late for some security applications.

```
# ./strcpy_stack 123456789123456789123456789
test
strcpy done
*** stack smashing detected ***: ./strcpy_stack terminated
Aborted
#
```

**Fig 3 functional test succeeded in protecting variable in the stack**

```
# ./strcpy_malloc 123456789123456789123456789
test
strcpy done
#
```

**Fig 4 functional test failed in protecting variable in the heap**

### 4.2.2 SPEC2006

In the SPEC2006 benchmarks, we added *start_protect* API function before the secondary function called, and *end_protect* API function after the secondary function returned. We selected *printf*, *fprintf*, *memset*, *strcpy*, *sprintf* as unreliable secondary functions.

We choose bzip2, mcf, libquantum as our test benchmarks. We use the test level input data for these benchmarks. Results show that the PULP configuration overhead is almost negligible, the total overhead of the PULP is less than 0.1% of the total execution runtime. The reason is that PULP is implemented on the basis of hardware check and the selected secondary functions are not called very frequently. So the configuration overhead is low compared to overall execution time.

Figure 5 shows the execution time comparison of old RISC-V and PULP. Their execution time are almost the same.

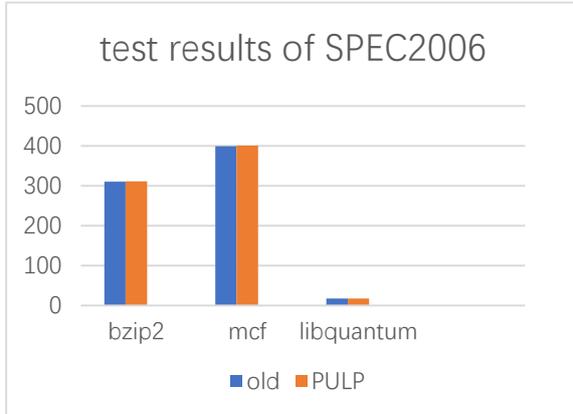

**Fig 5 the SPEC2006 execution seconds of OLD version RISC-V and PULP**

### 4.3 Hardware Cost Evaluation

Table 4 gives the hardware comparison results of old rocket chip and PULP. The area of PULP is bigger than the old rocket chip version by 31%. The cells of PULP is more than rocket chip by 2%. The power of PULP is more than rocket chip by 28%.

**Table 4 hardware comparison results of RISC-V old version and PULP**

|      | Area(um$^3$) | Cells | Power(W) |
|------|--------------|-------|----------|
| Old  | 53382        | 11648 | 3.543    |
| PULP | 69959        | 11885 | 4.540    |

## 5. Related works

### 5.1 Classic software measurements against Buffer Overflow

Stack canaries are special values stored in stack frames between the return address and local variables [17]. A contiguous stack buffer overflow would overwrite the stack canary, which is checked for intactness before the RETs of vulnerable functions.

Libsafe replaces the dynamic link library with new one [14]. Libsafe is a run-time solution that inserts wrapper code to estimate the maximum safe size for each destination buffer at the start of functions that are deemed to be vulnerable to buffer overflows. If the size of the data written to the destination buffer does not exceed the maximum safe size, then no return addresses can be overwritten.

LibsafeXP divides memory regions into global buffers, dynamically allocated heap buffers and stack buffers[4]. The global buffer's size and starting address are extracted from the symbol table section of ELF executable file, and dynamically allocated heap buffer's size and location information are tracked at run-time in the intercepted malloc family functions.

The stack protector implemented in GCC and clang adds an additional guard variable to each function's stack area[5]. The variable is initialized with a special value on function entry, and checked again on exit. Once the value has changed, the program aborts to prevent further damage. Adding these checks will lead to a little runtime overhead. The main drawback of stack protector is it can protect only variables in the stack.

### 5.2 Memory Protection Extensions (MPX)

Intel MPX [6] is a hardware accelerated memory corruption detection and prevention system. MPX requires program recompilation and library recompilation for full mitigation. The recompilation adds explicit instructions to the CPU for performing memory corruption checks. However, MPX has up to 4x overhead in RAM if the program has lots of pointers (trees, lists, graphs, etc).

### 5.3 Other In-process security techniques

Dune [7] is a system that provides applications with direct but safe access to hardware features such as ring protection, page tables, and tagged TLBs, while preserving the existing OS interfaces for processes. It consists of a small kernel module that initializes virtualization hardware and mediates interactions with the kernel, and a user-level library that helps applications manage privileged hardware features.

Shreds [8] is proposed to protect secret data (e.g., crypto keys and user passwords) and critical codes (e.g., private APIs and privileged functions). A shred can be viewed as a flexibly defined segment of a thread execution. Shreds offer in-process private memory without relying on separate page tables, nested paging, or even modified hardware based on ARM memory domains.

### 5.4 Comparison of in-process isolation techniques

In Table 5, we list the features of techniques above, and compare the performance overhead and shortcomings of these techniques.

**Table 5 features of recent isolation techniques**

|  | In-process security | | Memory bound check | Hardening c/c++ programs | wrapper functions checkout |
| --- | --- | --- | --- | --- | --- |
| techniques | shreds | Dune | MPX | Stack protector | LibsafeXP |
| characteristics | ARM DACR holds the access permissions for 16 domains, to isolate sensitive data | VT-x provides user programs with full access to protection hardware | dynamic detection and prevention of out of bound memory access | adds an additional guard variable to each function's stack area | Contains wrapper functions for buffer related functions in C standard library, enforces bounds checking |
| performance overhead | 4.67% for open-source applications | sandbox overhead is 2.9% for SPEC | 4x | low | 10% |
| shortcoming | use system call to overwrite the page protection. | cross privilege levels data sharing is not fast | runtime overhead is high | only protect stack variables | need additional instructions to check buffer bounds |

### 6. Conclusion

Based on hardware modification to the CPU architecture, PULP is an efficient user level in-process memory isolation system. Experimental results show that PULP could prevent in-process abuse attacks, with little overhead.

Better than most of software countermeasures of buffer overflow, such as Intel MPX, PULP can reduce software runtime overhead substantially. As long as PULP doesn't need additional instructions to check the memory access address, the runtime overhead of PULP is negligible.

Like Dune, PULP classifies user process into reliable primary functions and untrusted secondary functions, endowing different privileges to the two parts, realizing in-process data isolation, finally improving the security of user process.

Like shreds, PULP can realize in-process and thread-level isolation. Unlike shreds and Dune, PULP doesn't need kernel mode interference to achieve memory isolation. PULP doesn't need to step in kernel mode to change the hardware virtualization registers, so PULP can save much operating system runtime overhead, such as context switching overhead.

## References

[1] Paul Kocher, Daniel Genkin, Daniel Gruss, Werner Haas, Mike Hamburg, Moritz Lipp, Stefan Mangard, Thomas Prescher, Michael Schwarz, Yuval Yarom "Spectre Attacks: Exploiting Speculative Execution. " arXiv:1801.01203 [cs.CR]

[2] Zitser, Misha, R. Lippmann, and T. Leek. "Testing static analysis tools using exploitable buffer overflows from open source code. " ACM Sigsoft Twelfth International Symposium on Foundations of Software Engineering ACM, 2004:97-106.

[3] Limer, Eric . "How Heartbleed Works: The Code Behind the Internet's Security Nightmare. " Retrieved January 15, 2018.

[4] Lin, Zhiqiang, B. Mao, and L. Xie. "LibsafeXP: A Practical and Transparent Tool for Run-time Buffer Overflow Preventions." Information Assurance Workshop IEEE Xplore, 2006:332-339.

[5] http://www.productive-cpp.com/hardening-cpp-programs-stack-protector/. Retrieved January 15, 2018.

[6] Otterstad, C. W. "A brief evaluation of Intel®MPX." Systems Conference IEEE, 2015:1-7.

[7] Belay, Adam, et al. "Dune: safe user-level access to privileged CPU features." Usenix Conference on Operating Systems Design and Implementation USENIX Association, 2012:335-348.

[8] Chen, Yaohui, et al. "Shreds: Fine-Grained Execution Units with Private Memory." Security and Privacy IEEE, 2016:56-71.

[9] Zhang, Liang, et al. "Analysis of SSL certificate reissues and revocations in the wake of heartbleed." Conference on Internet Measurement Conference ACM, 2014:489-502.

[12] https://en.wikipedia.org/wiki/Buffer_overflow. Retrieved January 15, 2018.

[13] "Cyberoam Security Advisory – Heartbleed Vulnerability in OpenSSL. " April 11, 2014. Retrieved January 15, 2018.

[14] Tsai, Timothy K., and N. Singh. "Libsafe: Transparent System-wide Protection Against Buffer Overflow Attacks." International Conference on Dependable Systems and Networks IEEE Computer Society, 2002:541.

[15] "CVE – CVE-2014-0160. " Cve.mitre.org. Retrieved January 15, 2018.

[16] "CWE – CWE-126: Buffer Over-read (2.6). " Cwe.mitre.org. February 18, 2014. Retrieved January 15, 2018.

[17] Dang, T. H. Y., Maniatis, P., Wagner, D. "The Performance Cost of Shadow Stacks and Stack Canaries." ACM Symposium on Information, Computer and Communications Security ACM, 2015:555-566.

[18] https://samate.nist.gov/SRD/index.php Retrieved January 15, 2014.

[19] "The RISC-V Instruction Set Manual ." Volume II: Privileged Architecture    Version 1.9.1

[20] "The RISC-V Instruction Set Manual."    Volume I: User-Level ISA    Version